\documentclass[apj]{emulateapj}

\newcommand{\FeI}{\ion{Fe}{1}}
\newcommand{\CaII}{\ion{Ca}{2}}

\newcommand{\aliemail}{ali@nso.edu}
\newcommand{\naemail}{na.deng@csun.edu}
\newcommand{\debriemail}{debiprasad.choudhary@csun.edu}
\newcommand{\haiminemail}{haimin@flare.njit.edu}
\newcommand{\chungemail}{cliu@bbso.njit.edu}

\newcommand{\carstenemail}{cdenker@aip.de}

\shorttitle{Flow Field Evolution of a Decaying Sunspot}
\shortauthors{Deng et al.}

\begin{document}

\title{Flow Field Evolution of a Decaying Sunspot}

\author{Na Deng and Debi Prasad Choudhary}
\affil{California State University Northridge, Physics
       and Astronomy Department, 18111 Nordhoff St., Northridge, CA 91330; \naemail, \debriemail}

\author{Alexandra Tritschler}
\affil{National Solar Observatory/Sacramento Peak\footnote{Operated by the        Association of Universities for Research in Astronomy, Inc. (AURA),        for the National Science Foundation},
       P.O.~Box 62, Sunspot, NM 88349, U.S.A.; \aliemail}

\author{Carsten Denker\altaffilmark{1}}
\affil{ Astrophysikalisches Institut Potsdam, An der Sternwarte 16,
        D-14482 Potsdam, Germany; \carstenemail}

\author{Chang Liu and Haimin Wang}
\affil{New Jersey Institute of Technology, Physics Department,
       Center for Solar-Terrestrial Research, 323 Martin Luther King Blvd, Newark,
       NJ 07102 \\
       Big Bear Solar Observatory, 40386 North Shore Lane, Big Bear
       City, CA 92314, U.S.A.; \chungemail, \haiminemail}

\altaffiltext{1}{New Jersey Institute of Technology, Physics Department,
                 Center for Solar-Terrestrial Research, 323 Martin Luther King Blvd, Newark,
                 NJ 07102, U.S.A.}

\begin{abstract}

We study the evolution of the flows and horizontal proper motions in and around a decaying follower sunspot based
on time sequences of two-dimensional spectroscopic observations in the visible
and white light imaging data obtained over six days from June~7 to~12, 2005.
During this time period the sunspot decayed gradually to a pore.
The spectroscopic observations were obtained with the Fabry-P\'{e}rot based Visible-Light Imaging
Magnetograph (VIM) in conjunction with the high-order adaptive optics (AO) system operated
at the 65\,cm vacuum reflector of the Big Bear Solar Observatory (BBSO).
We apply local correlation tracking (LCT) to the speckle reconstructed
time sequences of white-light images around 600\,nm to infer horizontal proper motions
while the Doppler shifts of the scanned \FeI\ line at 630.15\,nm are used
to calculate line-of-sight (LOS) velocities with sub-arcsecond resolution.
We find that the dividing line between radial inward and outward proper motions in the
inner and outer penumbra, respectively, survives the decay phase.
In particular the moat flow is still detectable after the penumbra disappeared.
Based on our observations three major processes removed flux from the sunspot:
(a) fragmentation of the umbra, (b) flux cancelation of moving magnetic features (MMFs; of the same polarity as the sunspot)
that encounter the leading opposite polarity network and plages areas,
and (c) flux transport by MMFs (of the same polarity as the sunspot) to the surrounding network and plage regions
that have the same polarity as the sunspot.
\end{abstract}


\section{Introduction}\label{sec:introduction}

The evolution of active regions has been extensively studied for
decades  \citep[e.g. ][and references therein]{solanki2003}.  While
much progress has been achieved in understanding of how flux is
transported from deep in the convection zone to emerge at the solar
surface in form of sunspots, a comprehensive picture of the
regularities underlying the decay process
\citep[e.g.][]{martinezpillet2002} is still missing. Typically, a
statistical approach is followed to characterize the decay phase by
changes in umbral and penumbral  brightness, magnetic flux, or area in
an attempt to find either a common (mean) decay law
\citep[e.g.][]{morenoinsertis+vazquez1988, petrovay+martinezpillet+vandrielgesztelyi1999}
or by studying the distribution of the decay rates
\citep[e.g.][]{martinezpillet+morenoinsertis+vazquez1993}. However, this strategy  does only
evaluate the behavior of groups rather than of individual  sunspots
and does not give detailed insight into the individual processes that
might contribute to or trigger the decay process.

The decay phase of a sunspot can be initiated at any time
\citep{mcintosh1981} and proceeds differently for individual
sunspots. \citet{chapman+etal2003} analyzed the decay phase of many
sunspots and found a strong correlation between the sunspot
photometric decay rate, the total sunspot area and the umbral to total
area ratio. By carefully measuring the  magnetic flux,
\citet{zwaan1992} detected no significant increase in the total
magnetic flux of the sunspot after the penumbra is
formed. Furthermore, \citet{leka+skumanich1998} observed that the
umbral flux does not decrease during the formation of a
penumbra. These findings suggest that the formation and decay of
penumbrae is not  immediately related to changes in the sunspot
magnetic flux, but rather a phenomenon related to a change  in the
local flow field in and around the flux concentration. Differences
between the flow field around the sunspot and that around
the umbral fragments and pores thus might allow to determine critical
conditions necessary for both penumbral formation and decay.

The most prominent flow spectroscopically measured  in a sunspot is
the Evershed flow \citep{evershed1909},  which reveals itself in form
of shifts and  asymmetries of solar spectral lines and which is
interpreted as an outward mass flow in the photospheric layers of
penumbrae.  The Evershed effect is intrinsically coupled to the
presence of a penumbra and ceases at the immediate outer penumbral
border.  \citet{leka+skumanich1998} and \citet{yang+etal2003} detected
an Evershed flow as soon as the magnetic field lines reached a critical
inclination and organized filamentary structures became visible.  The
Evershed flow itself varies quasi-periodically in time
\citep{shine+etal1994, rimmele1994, rouppe2003,
cabrera+etal2006a}, which is predominately caused by velocity
patches that propagate along penumbral filaments from the middle penumbra to the
outer penumbra, the so-called Evershed clouds (ECs) which were first observed and
named by \citet{shine+etal1994}.  In their most recent study,
\citet{cabrera+etal2007} identify two classes of ECs, those that
vanish directly at the outer penumbral boundary and those that cross
the visible border and enter the moat flow. Once detached from the
penumbra, these ECs survive $\sim$14\,min before they disappear in
close proximity to the spot ($\sim$2\arcsec).  The outward migration
of the ECs could give a natural explanation for the existence of a dividing
line within the penumbra separating features that move inwards
in the inner penumbra and outwards in the outer penumbra.  This
dividing line appears typically in horizontal proper motion maps derived from either local
correlation tracking \citep[e.g.][]{november1986, november+etal1986}
or pattern recognition \citep[e.g.][]{sobotka+brandt+simon1999,
bovelet+wiehr2003}. \citet{molownyhoras1994} was the first to
identify such a ring-like dividing area, featuring near-zero
horizontal speed  and constant velocity divergence.

The sunspot moat is a zone that surrounds most mature sunspots and
some large sunspot pores.  The moat region is characterized by a
radially directed outward mass flow.  Although the moat is free of
stationary magnetic fields, time sequences of magnetograms show moving
magnetic features (MMFs), i.e., small-scale magnetic flux concentrations of
mixed polarity,  that move along with the moat flow in radial
direction away from the sunspot  \citep[]{sheeley1969,
harvey+harvey1973, brickhouse+labonte1988, yurchyshyn+wang2001, hagenaar+shine2005b,
kubo+shimizu+tsuneta2007, choudhary+bala2007}. There is increasing evidence that the
Evershed flow,  the moat flow, and MMFs are closely related to each other
\citep[e.g.][]{sainzdalda+martinezpillet2005, ravindra2006,
cabrera+etal2006a}. In a case study
\cite{vargasdominguez+etal2007} find that for a complex active region
the existence of the moat flow is coupled to the presence of
penumbral filaments. Hence, the moat should neither
be found around individual pores or umbral fragments after
the penumbra disappeared. However, observational results differ considerably
depending on what type of sunspot and pores have been observed
and what techniques have been used
\citep[see e.g.][]{wang+zirin1992, denker1998, yang+etal2003}.

In this paper, we study the evolution of line-of-sight (LOS) flows and horizontal proper motions in and around a
decaying sunspot. Our investigation is based on sequences of
high-spatial resolution observations obtained  over six days with new
post-focus instrumentation at the Big Bear Solar Observatory (BBSO)
\citep{denker+etal2005, denker+tritschler2005, denker+etal2007a}. Both
the LOS velocities measured  from two-dimensional spectroscopic
observations and horizontal proper motions derived with local correlation tracking (LCT) technique based on speckle reconstructed images are analyzed.  In \S\ref{sec:observation}
we overview the two-dimensional spectroscopic and speckle imaging
observations. The data reduction is described  in \S\ref{sec:reduction}
and the results are presented in \S\ref{sec:results}.
We discuss the results and offer our conclusions in \S\ref{sec:discussion}.

\section{Observations}\label{sec:observation}

\begin{figure}[t]
  \epsscale{1.}
  \plotone{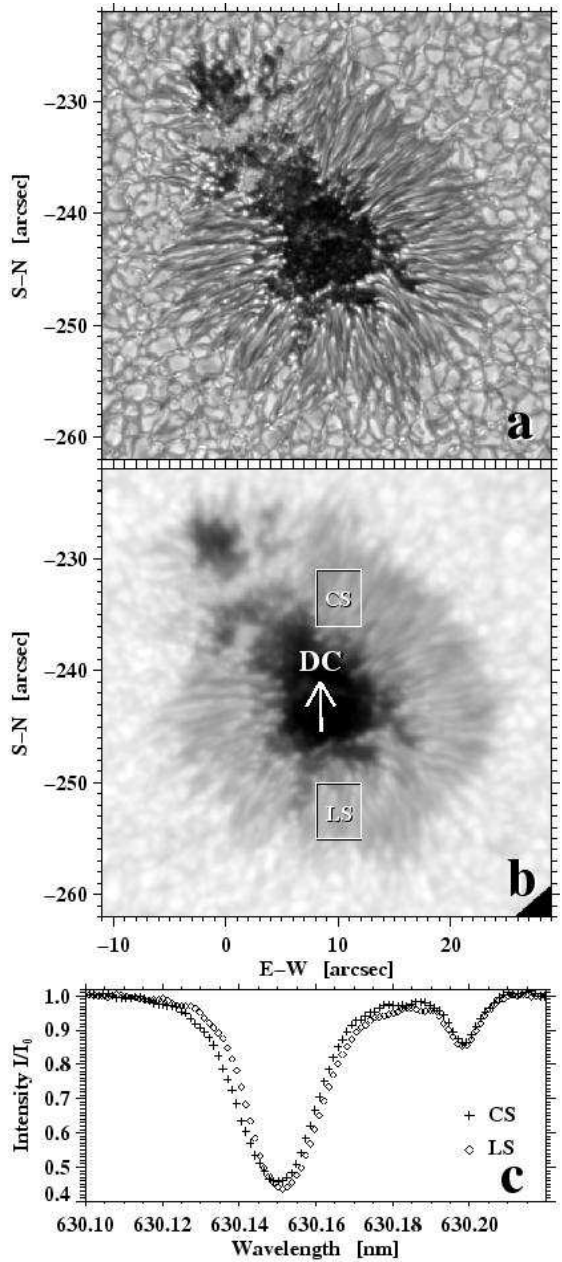}
  \caption{\textsl{a}: Speckle reconstructed white-light image of the follower sunspot in
           active region NOAA\,10773 on 2005 June\,7. \textsl{b}: Corresponding continuum
           intensity image observed with the two-dimensional spectrometer. The white arrow points towards
           disk center (DC). \textsl{c}: Sample line profiles of the two-dimensional spectroscopy
           averaged over small areas in the center-side (CS) and limb-side (LS)
           penumbra, which are indicated by white boxes in \textsl{b}.}
  \label{FIG3.1}
\end{figure}

The observations were performed with BBSO's 65\,cm vacuum reflector
in conjunction with the high-order adaptive optics (AO) system \citep{rimmele+etal2004,
denker+etal2007a}. We observed the isolated follower spot of active
region NOAA\,10773 during its disk passage
on five days in summer 2005: June\,7 ($\mu \approx
0.96$), June\,8 ($\mu \approx 0.95$), June\,10 ($\mu \approx 0.77$), June\,11
($\mu \approx 0.61$), and June\,12 ($\mu \approx 0.43$). Two-dimensional
spectroscopic scans of the photospheric spectral line \FeI\ 630.15\,nm
were obtained with the Visible-light Imaging Magnetograph (VIM) system,
in which a single Fabry-P\'erot etalon is the core element \citep{denker+tritschler2005}.
VIM has a bandpass of 8\,pm. A 0.3\,nm wide prefilter centered on the \FeI\ 630.15\,nm line
was used before the Fabry-P\'erot etalon. The \FeI\ line was sampled
at 101 equidistant wavelength points with a step size of 1.2\,pm corresponding
to a scanned wavelength interval of 0.12\,nm. The exposure time at each wavelength
point was 80\,ms. The cadence of each scan was about 18\,s. A
total of 100 scans can be obtained in 30\,min, which was the typical time range
of an observing run. The spectroscopic data have a field-of-view (FOV) of about $74\arcsec
\times 74\arcsec$. The 1k$\times$1k CCD camera was operated in a $2 \times
2$\,pixel binning mode. Thus, the image scale was about 0.14\arcsec\,pixel$^{-1}$. The spatial resolution achieved by the spectroscopic scans is estimated at about 0.5\arcsec\ with power spectrum analysis.

The spectroscopic observations are supported by quasi-synchronous bursts
of short-exposure (10\,ms) white-light images at 600\,$\pm$\,5\,nm
recorded with a fast CCD camera system for image restoration. Every 30\,s, a total of
200 frames was captured at 15 frames\,s$^{-1}$, from which the
best 100 frames were automatically selected and combined to create one restored
image using a speckle masking algorithm \citep{denker+etal2005}. The
cadence of the reconstructed white-light images is thus 30\,s. A total of
60 reconstructed images were obtained in a 30\,min observing run. The FOV is
about $70\arcsec \times 70\arcsec$ and the image scale is about
0\farcs07\,pixel\,$^{-1}$. The spatial resolution of the reconstructed images is diffraction limited ( $\approx$ 0.25\arcsec) by the 65~cm reflector of the BBSO if we use $1.22\lambda/D$ as the measure of the diffraction limit. Each observing run typically lasted 30\,min.
Further details of the observations are given by
\citet{denker+etal2007a,denker+etal2007b}, who provide a
thorough description of the high-resolution observations and a performance
evaluation of the AO system.

Figure~\ref{FIG3.1} shows the sunspot observed on June\,7 in a speckle
reconstructed white-light image (panel~\textsl{a}) and the corresponding continuum
intensity image obtained with VIM (panel~\textsl{b}). Panel~\textsl{c} shows two
sample line profiles averaged over a small area of
limb-side (LS) and center-side (CS) penumbra, which are indicated in panel~\textsl{b} by
white boxes. The two profiles are shifted w.r.t. each other
indicating the Evershed flow. The reconstructed image clearly shows the sunspot fine structure,
such as numerous umbral dots, the penumbral dark cores, and the
twist and writhe of the penumbral filaments. We also note that the penumbral
grains are brighter on the limb-side of the penumbra.

\begin{figure}[t]
  \epsscale{1.0}
  \plotone{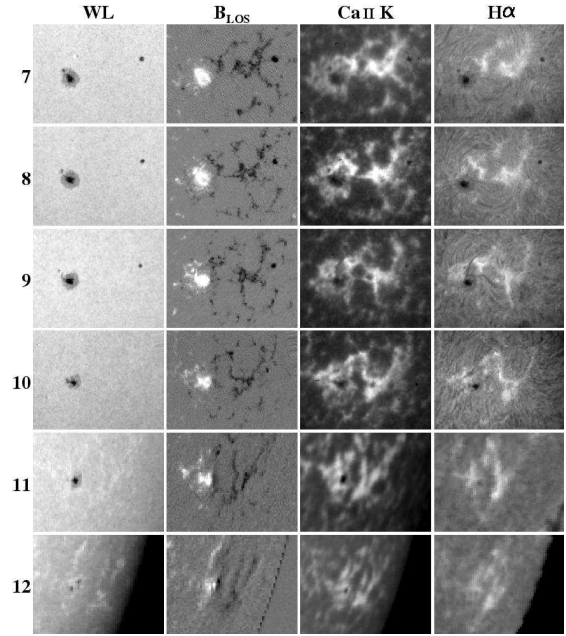}
  \caption{Evolution of active
           region NOAA\,10773 from 2005 June\,7 to June\,12. The
           columns (from \textsl{left} to \textsl{right})
           represent white-light (WL) images, LOS magnetic
           field $B_{LOS}$, and \CaII\,K and H$\alpha$ filtergrams.}
  \label{FIG3.2}
\end{figure}

NOAA\,10773 was also monitored by BBSO's 25\,cm refractor. These context observations include
vector magnetograms as well as H$\alpha$ and \CaII\,K filtergrams.
Figure~\ref{FIG3.2} shows the whole active region
and its evolution over six consecutive days based on
white-light information, LOS magnetograms, and \CaII\,K and H$\alpha$
filtergrams. BBSO's LOS magnetograms are produced by dividing the difference between the right- and left-hand circularly polarized Stokes components by their sum under weak field approximation. They suffer from the problem of Zeeman saturation when the field is stronger than typically 3000~G \citep{wang+etal1998,spirock2005}. We employ data from other observatories on some days because no synoptic BBSO data were available. For example, the white-light and magnetogram data on June\,9, 11, and 12 are from the
Michelson Doppler Imager \citep[MDI,][]{scherrer+etal1995} on board the Solar and
Heliospheric Observatory. The \CaII\,K and H$\alpha$ filtergrams on June\,11 and 12 are from the Mauna Loa Solar Observatory operated by the High Altitude Observatory. The active region was in the southern hemisphere (S15\arcdeg). The target was a relatively stable, round follower sunspot with positive magnetic
polarity located on the left side in the panels of Figure~\ref{FIG3.2}.
The leading spot on the right side was otherwise smaller in size. This configuration is an exception, since
the leading spot is typically more stable and larger in size. The bright
areas in \CaII\,K and H$\alpha$ filtergrams trace the plage regions that
are spatially correlated with magnetic fields of the supergranular network.

\section{Data Reduction}\label{sec:reduction}

We apply standard procedures such as dark subtraction and gain table correction
to the spectroscopic observations and white-light images. The speckle
reconstructions and the scans are carefully co-aligned with each other.
The white-light images were also co-aligned with the corresponding MDI
intensity images in order to acquire accurate locations and orientations
of the sunspot on different days. In order to remove the 5-minute oscillation
and residual seeing distortions a subsonic filter with a phase velocity
threshold of 4\,km\,s$^{-1}$ was applied to the time-series of white-light
speckle and spectroscopic data.

The LOS velocity is determined by a Fourier phase method
\citep[see][]{schmidt+stix+woehl1999}, which
uses the entire \FeI\, 630.15\,nm line profile and is very insensitive to noise. The telluric blend at 630.2~nm is not included in the calculation.
The determined shifts are converted to velocities by using
the Doppler formula. As a frame of reference we use the average
velocity of a small area in the darkest part of the umbra or
pore. The final calibrated dopplergrams follow the convention that redshifts are positive and blueshifts are negative. Thus, bright areas in the dopplergrams move away from the
observer, while dark areas move towards the observer. The \FeI\, 630.15\,nm
has a non-negligible Land\'e factor and is thus susceptible to
the presence of magnetic fields. The influence of the
magnetic field on the line shape, however, is symmetric with
respect to the line core and should not significantly influence
the velocity determination.

\begin{figure}[t]
  \epsscale{1.0}
  \plotone{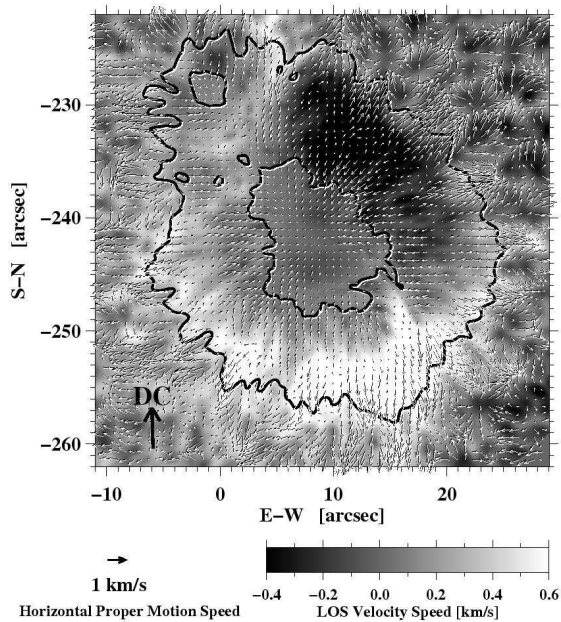}
  \caption{Combination of LOS dopplergram and LCT horizontal proper motions of the observed sunspot
           (NOAA~10773) on 2005 June\,7. Arrows show a 30-minute average of
           LCT proper motion maps derived from
           speckle-restored images. The gray-scale background image represents the
           LOS velocity, which is the corresponding 30-minute
           average of the dopplergrams. Redshifts are positive and correspond to
           bright areas. The black contours outline the umbral and penumbral boundaries.}
  \label{FIG3.3}
\end{figure}

\begin{figure*}[th]
  \epsscale{1.1}
  \plotone{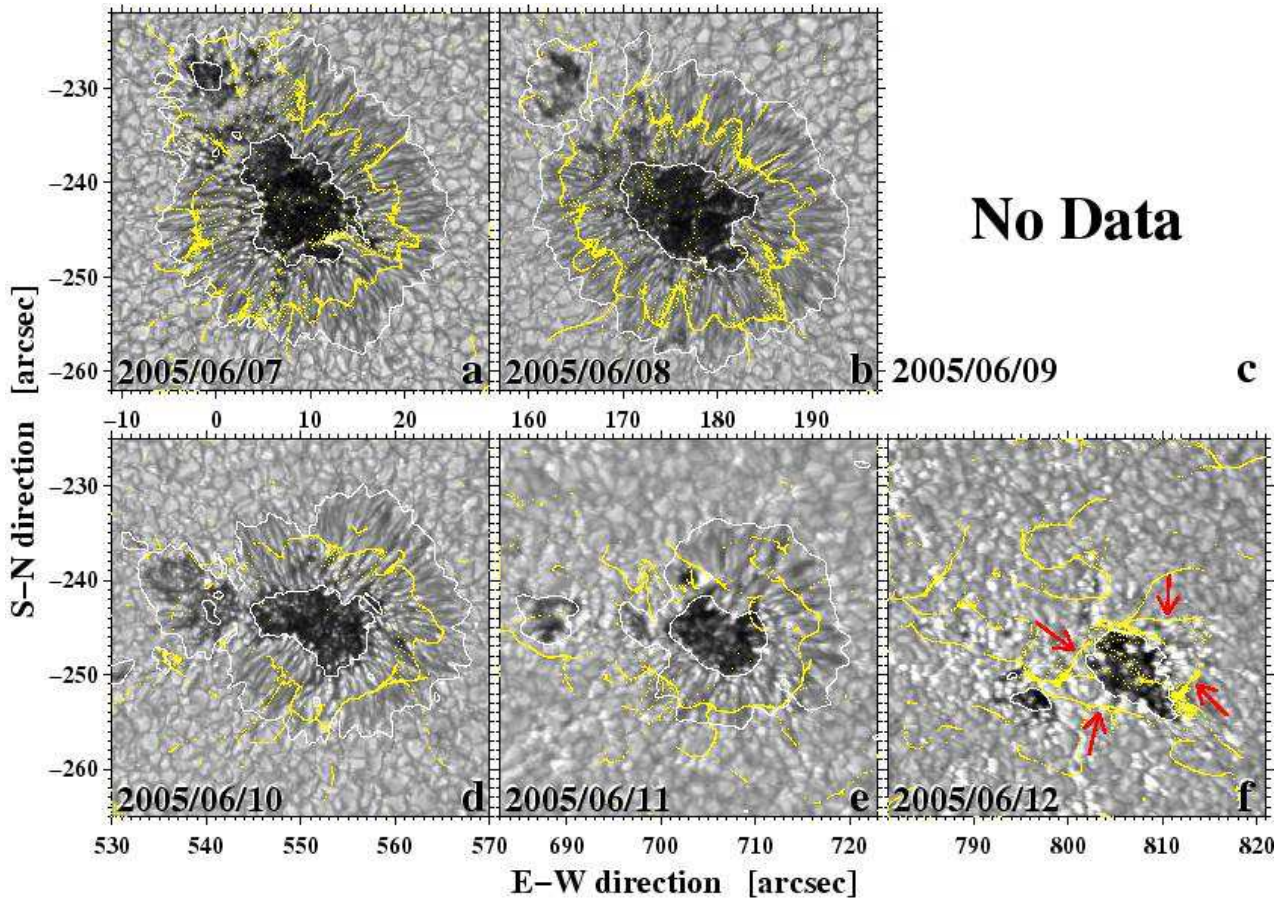}
  \caption{Sunspot evolution over six days during its decaying phase.
           White contour lines mark the umbral and penumbral boundaries
           determined from 60\,\% and 90\,\% of the quiet
           Sun intensity, respectively. The yellow corks trace the dividing
           lines between the inward
           and the outward proper motions in the penumbra and around the residual pore. The red arrows in panel \textsl{f} point to the broken and distorted dividing line around the pore.}
  \label{FIG3.4}
\end{figure*}

The ANA implementation (\url{http://ana.lmsal.com/ana/}) of the LCT algorithm \citep{november1986, november+etal1986} was applied to each of the individual 30-minute observing sequences of speckle
reconstructions. A Gaussian shape apodization window with a FWHM of 1.4\arcsec (corresponding to the mean size of the granulation) was used in the LCT technique. Since each sequence consists of 60 reconstructions, this resulted
in 59 individual LCT proper motion maps. The individual proper motion maps are averaged to
achieve a significant reduction of seeing induced random noise.
Furthermore, we use inverse cork maps to visualize sources of divergence in a LCT proper motion
map \citep[see e.g. ][]{molownyhoras1994, denker1998}. A total of 22,500 corks
(artificial tracer particles) were distributed evenly
across the average LCT proper motion map. Their movement is traced for 10\,h
backwards in time. The corks concentrate at positions where the proper motion shows strong and persistent divergence.
The resulting cork map traces the location of the dividing line between
inward and outward horizontal proper motions in the sunspot penumbra.
It should be mentioned that all the quantities and figures presented in this paper, unless otherwise noted, have been corrected for perspective foreshortening.

To demonstrate the outcome of the described methodology Figure \ref{FIG3.3}
displays the average LCT proper motion map superposed on the LOS dopplergram
averaged over the same time period for 2005 June\,7.
It clearly shows the Evershed effect in the
background LOS dopplergram, and visualizes the inward motion in the inner penumbra
and outward motion in the outer penumbra as indicated by the arrows showing the horizontal proper motion.
The outward proper motion exceeds the outer penumbral boundary by 3\arcsec\ to
5\arcsec\, while the LOS Evershed flow ceases more or less abruptly at the visible
boundary of the outer penumbra. Moreover, we detect diverging and converging trends
in the radial outward proper motions outside the sunspot.

\section{Results}\label{sec:results}

Figure~\ref{FIG3.4} illustrates the result of the cork analysis
based on the reconstructed white-light images for all observing days.
The white contours mark the umbral and penumbral boundaries as
derived from intensity thresholds set to 60\,\% and 90\,\% of the quiet
Sun intensity, respectively. The yellow dots indicate the corks that are
swept to regions where the horizontal proper motion is not divergence free.
Figure~\ref{FIG3.4} clearly reveals the dividing lines, which separate the inward
and outward proper motions in the penumbra. The dividing lines always reside inside
the penumbra. They tend to be more symmetric when the sunspot penumbra itself has a more
symmetric shape (see panel \textsl{b} in Figure~\ref{FIG3.4}). They lose
symmetry when the sunspot penumbra becomes less symmetric (see the other panels
in Figure~\ref{FIG3.4}). Interestingly, even after the penumbra completely
disappeared, there is still a distinct dividing
line around the remaining pore, although, incomplete
and disrupted at some locations (see panel \textsl{f} in Figure~\ref{FIG3.4}).

\begin{figure}[t]
  \epsscale{1.}
  \plotone{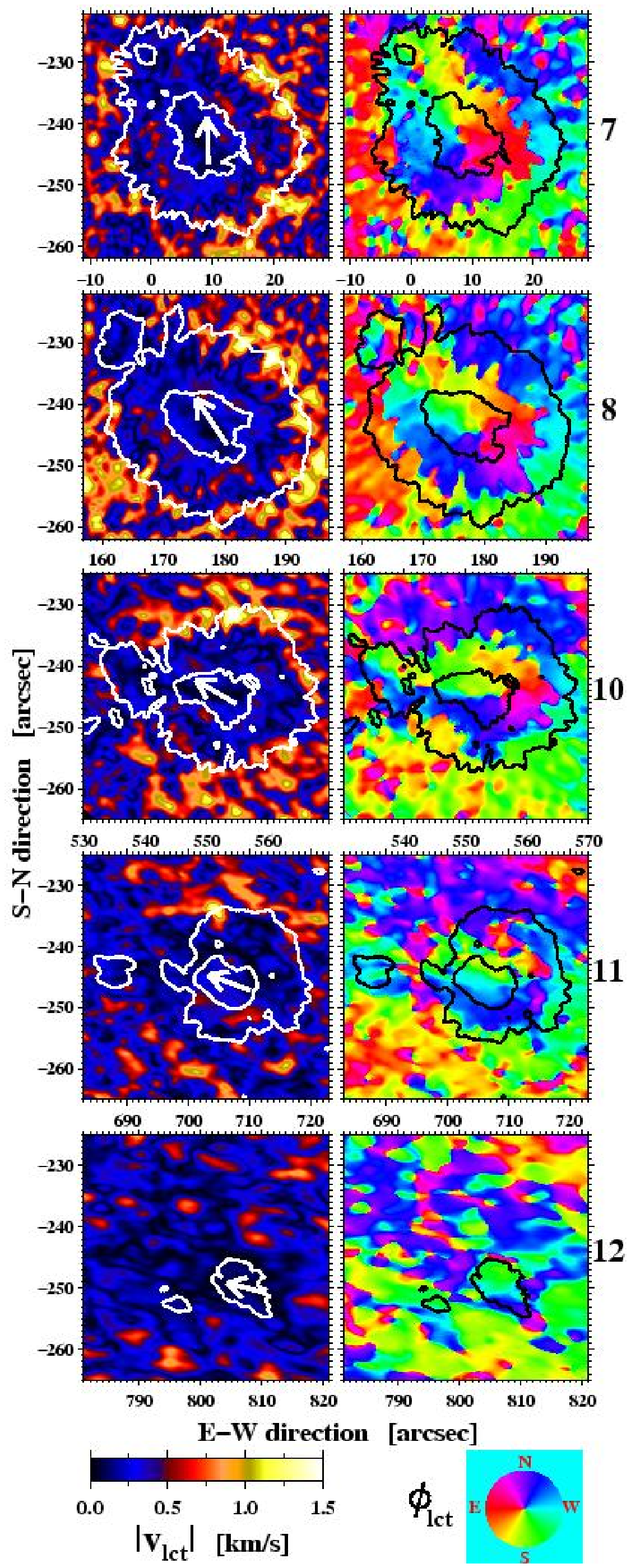}
   \caption{Decomposition of the 30\,min average horizontal proper motions derived from LCT into speed
    ($|v_{\rm LCT}|$, \textsl{left}) and azimuth maps ($\phi_{\rm LCT}$, \textsl{right}). The white or black contours outline the
   umbral and penumbral boundaries of the sunspot. The white arrows in the
    umbrae point towards the direction of disk center. The magnitude of the speed is
    indicated by the color scale and the azimuth angle is indicated by the color-coded disk. }
  \label{FIG3.6}
\end{figure}
\begin{figure*}[t]
  \epsscale{1.1}
  \plotone{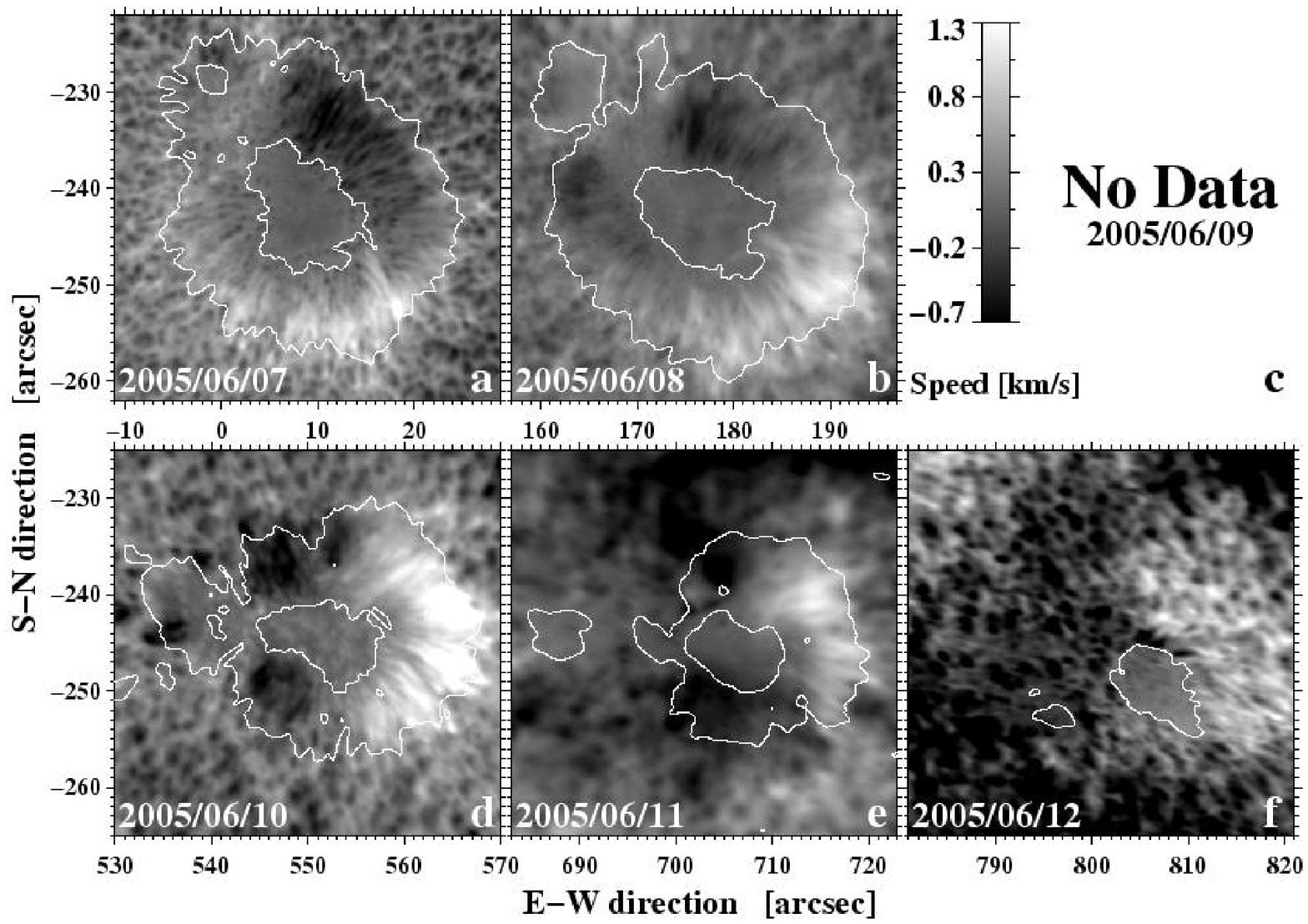}
  \caption{LOS dopplergram series corresponding to Figure~\ref{FIG3.4}
           showing the Evershed flow in the
           sunspot. Redshifts are positive corresponding to bright areas. The gray
           scale in Panel c represents the magnitude of LOS velocity.}
  \label{FIG3.7}
\end{figure*}

The average LCT proper motion vectors (see Figure~\ref{FIG3.3}) are decomposed
into two components, i.e., magnitude $|v_{\rm LCT}|$ and azimuthal
direction $\phi_{\rm LCT}$. They are displayed for each pixel in
Figure~\ref{FIG3.6}.  From June\,7 to June\,11, when the sunspot has a
penumbra, the $\phi_{\rm LCT}$ component shows a radial inward motion
in the inner sunspot (including the umbra and the inner part of
penumbra) and a radial outward motion in the outer sunspot. The
outward motion extends beyond the visible boundary of the outer
penumbra for up to 5\arcsec.  In other words, a 5\arcsec\-wide annular
zone showing a radial outward proper motion (collar flow) surrounds the penumbra (or the
sunspot).  The $|v_{\rm LCT}|$ component shows that there are two
ring-like proper motion structures with large $|v_{\rm LCT}|$ values. One is
surrounding the penumbra-umbra interface corresponding to the inward
motion of penumbral grains, which has relatively small magnitude (with
average speeds of about 0.4\,km\,s$^{-1}$ and a maximum speed reaching
up to 0.8\,km\,s$^{-1}$). Its width is about 3\arcsec\
($\sim$2200\,km). The other is surrounding the outer penumbra and
harbors the high speed portion of the outward motion of penumbral
structures and the immediate part of the moat zone. There the $|v_{\rm LCT}|$ component
exhibits a larger magnitude with an average speed of about
0.8\,km\,s$^{-1}$ and a maximum speed reaching up to
1.4\,km\,s$^{-1}$. Its width is about 5\arcsec\
($\sim$3700\,km). \citet{november+etal1987} were the first to detect
a radial, large speed collar flow of granules from the penumbra into the immediate
surroundings from the flow field around sunspots
and named it ``annulus''. The ring-like motion patterns lose their symmetry when the
sunspot decays. Both inner and outer ring seem to be shifted towards
the limb-side for about 2\arcsec\ or 1500\,km. As can be seen in
Figure~\ref{FIG3.6}, the rings at the center-side occur at the
penumbra-granulation or penumbra-umbra boundary, while the rings at
the limb-side occur outside corresponding borders.

Concentrating on the residual pore on June\,12, we see from the $\phi_{\rm LCT}$
component that there is still a confined region ($\sim$2\arcsec\ to 3\arcsec)
around the pore where we detect a horizontal proper motion towards the pore, which is also indicated by the
dividing line seen in Figure~\ref{FIG3.4}. Beyond the dividing line
$\phi_{\rm LCT}$ indicates a motion that is directed away from the pore.
The $|v_{\rm LCT}|$ component shows that both the inward and outward motion in and
around the pore are not as strong as those in and around the sunspot.

The LOS dopplergrams are displayed in Figure~\ref{FIG3.7} for each day
corresponding to the arrangement in Figure~\ref{FIG3.4}. The dopplergrams
illustrate not only the dominant Evershed effect, but also the fine structure in
the LOS velocity of the whole sunspot. This includes an indication of
upflows (dark blueshifts) in umbral dots, the presence of penumbral flow channels,
and upflows in penumbral grains, which are more prominent
on the limb-side of the penumbra with a bright redshifted background. Furthermore,
the dopplergrams corroborate the results derived from LCT proper motion maps and the cork analysis.
In particular, the LOS flow field around the pore on June\,12 strongly suggests that the moat flow
is still present: on the limb side and center side of the pore we observe
a large patch of redshifts and blueshifts, respectively. At the large viewing angle
($\mu = 0.43$), where also higher layers of the solar atmosphere
are probed, the moat flow component adds significantly
to the convective flow pattern in such a way that it increases the LOS
component of the granular upflows on the center side but overcompensates
the LOS component of the granular upflows limbwards of the pore. It is noticed that the dopplergrams for June 11 and 12 show much stronger LOS velocities far from the spot or pore than observations nearer disk center (e.g., June 7). The reason for that could be the convection of the supergranulation that contributes more to the LOS velocity towards the limb.

From Figures~\ref{FIG3.2} and \ref{FIG3.4}, and by watching long-term intensity
and magnetogram movies from MDI, our visual impression is that magnetic flux was removed
from the sunspot by the following processes: fragmentation of the
spot in the north-eastern part and by MMFs.
To the west, the flux transported by those MMFs that have the same polarity as the spot
was promptly canceled by the leading negative flux
when the MMFs reached the outer edge of the moat region. To the east
where the network and plage showed much less opposite
polarity flux, substantial amount of flux
was transported by MMFs of the same polarity as the sunspot into the surrounding plage region of the same polarity.
Fragmentation, flux cancelation, and transport of flux to the
network and plage region were likely responsible for the decay of the sunspot.
On June\,12, the penumbra completely disappeared and only a pair
of pores remained. In addition, the sunspot had rotated counterclockwise by about
45\arcdeg\ during the five days from June\,7 to June\,11.

We also noticed that the relative brightness of both penumbra and umbra
increased when the sunspot moved from disk center to the limb. Moreover, this
brightness increase is faster in the umbra than in the penumbra. The decay of
the sunspot could be responsible for the overall increase of the sunspot
brightness. Another possibility is that the sunspot is relatively brighter from
the lower to the upper photosphere, which suggests that the temperature along
magnetic flux tube increases faster than that in weak or non-magnetic plasmas. This behavior of the thermal stratification agrees well with the cool umbral model of \citet{collados+etal1994}.

\begin{figure}[t]
  \epsscale{1.1}\vspace{0.5cm}
  \plotone{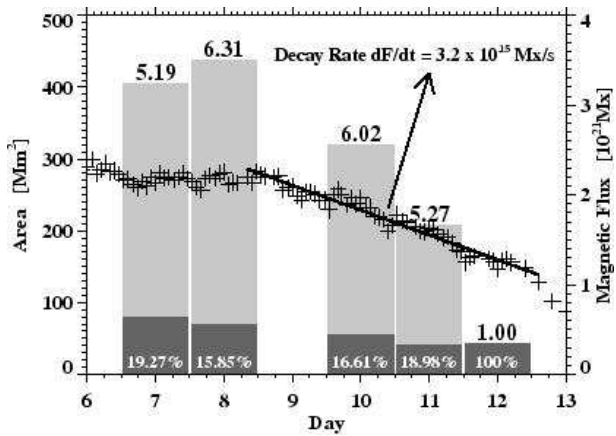}
  \caption{Evolution of sunspot area and magnetic flux.
           The light gray bars represent the penumbral area,
           dark gray bars correspond to the umbral area. Their sum (i.e., the whole bar)
           relates to the total area of the sunspot. The series of plus signs shows
           the evolution of the magnetic flux of the sunspot.
           A linear fit is applied to the decaying
           phase from June\,9 to 12. The slope provides the overall
           decay rate of $3.2 \times 10^{15}$\,Mx\,s$^{-1}$.}
  \label{FIG3.5}
\end{figure}

To illustrate the decay process of the sunspot more quantitatively we
determined the change of the total sunspot area and the magnetic flux with time.
The result is shown in Fig.~\ref{FIG3.5}. The light and dark gray
vertical bars represent the penumbral and umbral areas, respectively. Their sum
corresponds to the total area of the sunspot. The areas were calculated using
the high-resolution data based on the enclosed areas shown by contours in
Figure~\ref{FIG3.4}. The total-to-umbral area ratio
($r_{A} = A_{\rm tot}/A_{U}$) and the percentage of the umbral to total sunspot
area for each day were also calculated and annotated accordingly on the top and at
the bottom of each bar in Figure~\ref{FIG3.5}, respectively. The published
results of $r_{A}$ range from 4.0 to 6.3, but are mostly distributed between
5.0 and 6.0 \citep[for a summary see][]{solanki2003}. The result from this study
shows that $r_{A}$ ranges from 5.19 to 6.31, which is in agreement with
previous measurements. The percentage of the umbral to total sunspot area
increased from 15.85\,\% to 18.98\,\% from June\,8 to June\,11, which indicates that the
sunspot penumbra decays faster than the umbra. The plus signs in
Figure~\ref{FIG3.5} show the evolution of the magnetic flux in the sunspot ($F
= \int B \cdot dS$) calculated from MDI magnetograms. The projection effect of
the area $S$ has been corrected and the magnetic fields $B$ were
partially corrected assuming that the field lines are horizontal in the penumbra and vertical in the umbra. From
June~9 to 12, the sunspot flux decayed with an almost constant decay rate of
$3.2 \times 10^{15}$\,Mx\,s$^{-1}$.

\section{Discussion and Conclusion}\label{sec:discussion}

A photometric and spectroscopic high-resolution study has been presented
showing the flow fields in a decaying, isolated, quasi-round sunspot covering
five days in June\,2005. Based on horizontal proper motions and LOS flow fields determined
from a LCT analysis and spectroscopic observations, respectively,
we confirm the existence of inward motions in the inner penumbra and
outward motions in the outer penumbra separated by a dividing line
located inside the penumbra \citep[see e.g.][]{zirin+wang1989, wang+zirin1992,
molownyhoras1994, denker1998, sobotka+brandt+simon1999, bovelet+wiehr2003}.
For the first time, we demonstrate that the dividing line survives the
decay process during which the sunspot lost its penumbra
and fragmented. Most interestingly, we find that the moat flow in the periphery of the
umbral core, i.e the residual pore, is still detectable after the penumbra has disappeared.
The outward motion, however, is not in the immediate surroundings of
the pore but separated by an annular inward motion, and is
much weaker than the one we find around the sunspot. We argue that the
residual pore retained the flow pattern bequeathed from its parent sunspot, since sunspots and pores have historic properties, i.e., they tend to continue their past physical properties, as evidenced by the fact that large pores can be larger than small sunspots \citep{solanki2003}.
The existence of a moat flow around umbral fragments or
individual pores is still debated in the literature
\citep[e.g. ][]{wang+zirin1992, yang+etal2003, vargasdominguez+etal2007}.
Generally, it is believed that the large-scale organized outflow pattern of
the moat around sunspots is driven by a temperature excess that builds
up beneath because normal convective energy transport is inhibited.
Furthermore, results from time-distance helioseismology corroborate
the existence of large-scale vortex cells and
downflows beneath sunspots \citep[e.g][]{duvall+etal1996, gizon+duvall+larsen2000,
zhao+kosovichev+duvall2001}. Already \citet{parker1992} suggested that individual
emerging flux bundles are collared by downdraft vortex rings.
Motivated by \cite{bovelet+wiehr2003} who put these ideas together illustrated
in a sketch (see their Figure~13) we explain our observation of
a collar flow directed towards the umbra after the penumbra has disappeared
as the signature of that downflow vortex cell, which is adjacent to an outer upflow and diverging vortex roll.

The inward proper motions in the inner penumbra can be very well explained in the framework of the
moving-tube scenario by \cite{schlichenmaier+jahn+schmidt1998a}
mimicking the uncombed penumbra \citep{solanki+montavon1993}.
In this model the filamentary structure of the penumbra originates
from the dynamics of ascending flux tubes embedded in a static background
field. During the ascending phase a systematic flow develops along the flux tube,
the Evershed flow. In particular penumbral grains are interpreted as the upstream
footpoints of the flux tubes. As the flux tubes rise their cross-section
with the $\tau=1$-level migrates inwards giving the
visual impression that the penumbral grains move towards the umbra.
Many observations have provided support for this model like most recently
\citet{rimmele+marino2006}. The outward migration detected in the outer
penumbra might be explained by the moving-tube model if the photospheric sea-serpent structure developed from the penumbral flux tubes simulated by \citet{schlichenmaier2002} occurs in the real Sun. We speculate that the outward flow which extends beyond the visible boundary of the sunspot penumbra is related to the propagation of ECs \citep[see][]{cabrera+etal2007}
but this needs still to be verified.

The observed sunspot was a peculiar follower in the sense that it was
larger, more regular and more stable than its leading spot. \citet{martinezpillet2002}
reviewed several aspects related to the decay of leading and following sunspots
and summarized the current state of knowledge of the physical mechanisms that
are involved in sunspot evolution. In general, a long period of stability is
rarely found in followers, they decay very fast, typically
within several days or less after their formation. Following spots usually
possess irregular shapes and have incomplete penumbrae. Only 3\,\% of the observed
followers develop a round shape and appear in the so-called $\alpha$ configuration
\citep{bray+loughhead1964}. It is much more likely for the preceding sunspot to
develop such a configuration which then can last from days to
weeks. Hence, the observations presented in this study provide a rare
exception of the aforementioned rules. We find that during the decay
phase the percentage of the umbral to total sunspot area
increased from 15.9\,\% to 19.0\,\%  indicating that the penumbra decays
faster than the umbra. Although the flux decay rate was almost constant
($3.2 \times 10^{15}$\,Mx\,s$^{-1}$), the decay process was not uniform.
We observed that the decay proceeded by the following mechanisms:
fragmentation of the sunspot, flux cancelation of MMFs (of the same polarity as the sunspot)
that encounter the leading opposite polarity network and plages areas, and flux transport by MMFs (of the same polarity as the sunspot)
to the network and plage regions that have the same polarity of the sunspot.

In order to reconcile the discrepancies found in the large-scale flow patterns
surrounding sunspots, umbral fragments, individual pores that never developed
a penumbra (innate pores), and individual pores that remained from a decayed sunspot (residual pores) not only better statistics is needed. To fully understand
the connection between the flow patterns and the evolution of the magnetic structures
high-spatial resolution spectropolarimetric observations as provided
by HINODE must complete spectroscopic and imaging observations like those
presented in this investigation.

\acknowledgments
We would like to thank John Thomas, Valantina Abramenco, and Dale E. Gary
for comments and discussions on the original manuscript. We also thank the referee for a number of valuable suggestions that help to improve the paper.
This research has made use of NASA's Astrophysics Data System (ADS).
This work was supported by NSF under grant ATM 03-42560, ATM 03-13591,
ATM 02-36945, ATM 05-48952, and MRI AST 00-79482 and by NASA under grant NAG 5-13661.

\end{document}